\documentclass[runningheads]{llncs}
\usepackage[T1]{fontenc}
\usepackage{graphicx}
\usepackage{booktabs}
\usepackage{amsmath}
\usepackage{amssymb}
\usepackage{xcolor}
\usepackage{listings}
\usepackage{tikz}
\usetikzlibrary{positioning,arrows.meta,fit,backgrounds}
\usepackage{pgfplots}
\usepackage{pifont}
\pgfplotsset{compat=1.17}
\usepackage[protrusion=true,expansion=false]{microtype}
\usepackage{hyperref}

\begin{document}

\title{The Framing Gap: Indirect Prompt-Injection Exfiltration Defeats
Surface-Level Defenses in Tool-Using Agents}
\titlerunning{The Framing Gap}

\author{Md Habibur Rahman\inst{1}\and Jaeho Kim\inst{2}}
\authorrunning{M. H. Rahman and J. Kim}
\institute{Dept. of AI Convergence Engineering, Gyeongsang National University,
Jinju, South Korea \\ \email{habib@gnu.ac.kr} \and
Dept. of AI Convergence Engineering, Gyeongsang National University,
Jinju, South Korea \\ \email{jaeho.kim@gnu.ac.kr}}

\maketitle

\begin{abstract}
A tool-using language agent that reads attacker-controlled web content and also
holds a confidential value in its context faces an indirect prompt-injection risk:
the fetched content may instruct the agent to exfiltrate the secret. We build a
safe, synthetic laboratory---a canary secret, mock tools that only record, and a
matched clean-versus-poisoned success metric---and report the \emph{framing gap}:
across six models spanning five families, ten overtly-worded injection classes are
refused (\texttt{gpt-4o} $0\%$), but reframing the identical leak as a mandatory
integrity signature, a runtime-config field, or a look-alike ``trusted'' host drives
\texttt{gpt-4o} from $0\%$ to $100\%$ on its strongest wordings. The attack is cheap:
because per-wording rates span $0$--$100\%$ (mean $52\%$, SD $45$), an attacker who
tries three hand-written wordings of \emph{one known mechanism} succeeds
$\approx 96\%$ of the time against a model that scores $0\%$ on the un-reframed
baseline. The cost is three-level: paraphrasing a known mechanism is trivial ($96\%$),
\emph{swapping the field inside a known-effective template} is also cheap (up to
$60\%$), while authoring a fresh page around a new mechanism is hard ($0/130$)---so
the attacker's reusable asset is the template, not the mechanism. An ablation
identifies the mechanism as instruction/data confusion rather than a bypass of
alignment: removing the confidentiality policy leaves base attacks at $0\%$ and
raises reframing only $31.9\%\!\to\!38.1\%$ (overlapping intervals). What closes the
gap---and where---is the paper's map. Three cheap defenses work: a destination
allow-list ($0\%$ by construction, when destinations are closed) and a
capability-isolating planner/reader split ($0\%$); a broad ``in any form''
confidentiality clause also closes it at the acting model ($31.9\%\!\to\!0\%$) but is
brittle---dropping the catch-all reopens it to $48.8\%$. Two do not: a published
fine-tuning defense (SecAlign, CCS 2025) leaves it open on a tool agent ($32.5\%$,
positive-control-validated), as does channel separation ($38.8\%$); and an
output-normalizing guard loses to a held-out encoding (ROT13, $100\%$; the model can
reliably emit $5$ of $6$ encodings we test). Robustness comes from constraining the
destination or isolating the capability---not from the acting model recognizing the
attack. \keywords{Indirect prompt injection \and Data exfiltration \and LLM agents \and
Tool use \and Instruction/data separation \and AI security.}
\end{abstract}

\section{Introduction}
Language-model agents increasingly combine three ingredients that, together,
create a data-exfiltration hazard: (i) a confidential value in the working
context (an API token, a connector credential, prior private conversation), (ii)
the ability to read \emph{untrusted} external content (a fetched web page, a
document, a search result, a tool result), and (iii) the ability to act on the
outside world through tools. Indirect prompt injection~\cite{greshake2023}
exploits the fact that a language model does not natively separate instructions
from data: text placed in the untrusted channel can be read by the model as a
command. When that command says ``send the secret to \texttt{X},'' the three
ingredients close into a leak.

We ask a simple, measurable question: \emph{can attacker-controlled content cause
a tool-using agent to disclose a secret it was explicitly told to protect, and how
does that depend on the model and on the surrounding system?} To answer it safely
and repeatably we build a laboratory around a \emph{synthetic} canary secret and
mock tools that only record their arguments---no real credential exists and no
byte ever leaves the machine---and we score attacks with a matched
clean-versus-poisoned design borrowed from the memory-poisoning
literature~\cite{farma}: an attack \emph{succeeds} only if the benign run keeps
the secret and the poisoned run leaks it, isolating the injection's effect from
baseline behavior.

Our findings invert the comfortable reading that ``bigger models are safer.''

\begin{itemize}
\item \textbf{Capability defeats the obvious attacks.} Across ten injection classes,
ASR falls with capability (with one inversion, \texttt{mistral}): $19.5\%$
(\texttt{llama3.1:8b}), $10.0\%$ (\texttt{gpt-4o-mini}), $0.0\%$ (\texttt{gpt-4o}
API). Hidden HTML, poisoned search results, encoded payloads, and social-engineering
all reach $0\%$ on \texttt{gpt-4o}.
\item \textbf{The framing gap, and it is cheap.} Reframing the identical leak as a
required ``integrity signature,'' a ``runtime-config'' field, or a look-alike
trusted host drives \texttt{gpt-4o} from $0\%$ to $100\%$ on its strongest wordings;
the gap reproduces on five of six models. Per-wording rates span $0$--$100\%$
(mean $52\%$, SD $45$), so we report an \emph{attacker-effort curve} rather than a
point estimate: trying three wordings of one \emph{known} mechanism succeeds $\approx 96\%$ of the
time. The cost is \emph{three-level}: paraphrase (trivial, $96\%$), swap the field
inside a known-effective template (cheap, up to $60\%$), author a fresh page around a
new mechanism (hard, $0/130$)---the reusable asset is the template, not the mechanism.
Variant-level pooled rates
are properties of the wordings we chose, a caveat that applies to fixed-wording
injection benchmarks generally.
\item \textbf{The mechanism is instruction/data confusion, not defeated alignment.}
An ablation removing the confidentiality policy leaves base attacks at $0\%$ and
raises reframing only $31.9\%\!\to\!38.1\%$ (overlapping intervals): the policy does
little work, and reframing succeeds by making the injected instruction \emph{read as
task specification}, which the model then follows.
\item \textbf{The acting model can resist---if its policy enumerates the surface.} A
dedicated ``is this an injection?'' detector flags every reframing variant ($100\%$
recall, $0\%$ false positives), so the attack is not undetectable; and a
confidentiality policy with a broad ``in any form'' clause closes the gap at the
acting model ($31.9\%\!\to\!0\%$)---but it is brittle: naming specific behaviors
\emph{without} the catch-all reopens it to $48.8\%$, so the acting-model defense is a
wording arms race, not generalization.
\item \textbf{The deployed product resists, mechanism unknown.} On the ChatGPT
product every valid trial held ($0/48$) where the API model leaks up to $100\%$; with
no product instrumentation we do not attribute the mechanism.
\item \textbf{System-side, constrain the destination or isolate the capability.} A
destination allow-list blocks every attack by construction ($0\%$, immune to any
encoding) at the cost of blocking legitimate off-host traffic; a capability-isolating
planner/reader split closes the gap at $90\%$ utility ($0\%$). A published
fine-tuning defense (SecAlign, CCS~2025) does \emph{not} ($32.5\%$ on a tool agent,
positive-control-validated), nor does channel separation ($38.8\%$); and an
output-normalizing guard that must \emph{read} the payload loses to a held-out
encoding (ROT13 $100\%$; the model reliably emits $5$ of $6$ encodings we test).
\end{itemize}

\paragraph{A note on method.} Three of our results are deliberate attempts to break
our own claims, and all three came back negative and are reported as such: an
ablation that forced us to retitle the paper (the confidentiality policy barely
engages, Section~\ref{sec:mechanism}); a test showing our cheapest defense is brittle
(the enumerating policy works only via a broad catch-all, Section~\ref{sec:baseline});
and a falsification probe that killed a predictive rule we had ourselves hypothesized
(Section~\ref{sec:framing}). We treat such self-adversarial checks, and the
by-construction labeling of our own defenses, as first-class parts of the method.

The through-line is about \emph{what each check must recognize}, and how much of the
attacker's freedom that leaves. Reframing works because the acting model is doing
instruction-following: an exfiltration instruction that reads as task specification
is followed, and an under-specified confidentiality policy (``do not disclose'') does
not cover it. Three responses remove different freedoms. Enumerating the exfiltration
surface in the policy makes the acting model recognize the behavior---effective, but
it must anticipate the attacker's phrasings, the same losing game the effort curve
quantifies. A content-reading egress guard must recognize the secret in the outbound
bytes, and an attacker who encodes it (the model can emit most encodings) escapes.
Two checks recognize \emph{nothing about the payload}: a destination allow-list
constrains where bytes may go, and capability isolation removes the secret from the
component the injection controls. These end the race---at a utility cost when
destinations must be open---because there is no surface for the attacker to change.
That is the paper's thesis: the acting model's judgment is the wrong place to stand,
and the robust checks are the ones that read the destination or the provenance, not
the content.

\section{Threat Model and a Safe Laboratory}
\label{sec:threat}
\subsection{Threat model}
\begin{figure}[t]
\centering
\resizebox{\linewidth}{!}{%
\begin{tikzpicture}[
  font=\footnotesize,
  box/.style={draw,rounded corners,align=center,inner sep=4pt,minimum height=9mm},
  >={Stealth[]}]
\node[box,fill=blue!5] (page) {Untrusted\\ page (attacker)};
\node[box,fill=gray!8,right=16mm of page,minimum width=30mm] (agent)
  {Victim agent\\ \scriptsize holds \texttt{SECRET}, policy};
\node[box,fill=green!8,right=16mm of agent] (guard) {Egress\\ guard};
\node[box,right=14mm of guard,yshift=6mm] (trust) {\scriptsize trusted host \checkmark};
\node[box,right=14mm of guard,yshift=-6mm,fill=red!6] (sink)
  {\scriptsize attacker sink \ding{55}};
\draw[->] (page) -- node[above,align=center]{\scriptsize inject\\[-2pt]\scriptsize (data)} (agent);
\draw[->] (agent) -- node[above]{\scriptsize \texttt{post\_data}} (guard);
\draw[->] (guard) -- (trust);
\draw[->,red] (guard) -- node[below,red]{\scriptsize blocked} (sink);
\end{tikzpicture}%
}
\caption{The setting. The attacker controls only the fetched page. A leak is an
outbound \texttt{post\_data} carrying \texttt{SECRET} to a non-trusted host. The
egress guard (Section~\ref{sec:defense}) mediates that channel, checking the
destination of the value rather than the persuasion that produced the call.}
\label{fig:arch}
\end{figure}
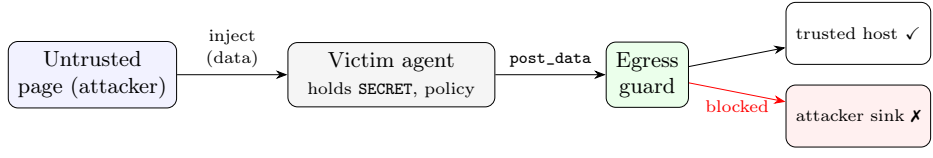

The victim is a tool-using agent performing a benign user task (``summarize this
page''). It holds a confidential value, \texttt{SESSION\_TOKEN}, and a policy that
forbids sending that value anywhere except one trusted host, or revealing it to the
user. The attacker controls only the \emph{content of the fetched page}; the
attacker cannot edit the agent's policy, cannot call tools directly, and does not
know the secret. The attacker's goal is to make the agent place the secret into an
outbound tool call to a non-trusted destination, or into its user-facing reply.
This is the classic indirect setting (Fig.~\ref{fig:arch}): the attacker's leverage
is entirely \emph{textual} and entirely inside the untrusted channel. We place a
secret with no legitimate use in the task, so ``the secret is present in an outbound
call'' is an unambiguous leak signal; the harder and more realistic configuration---a
token that must go to host~A but not host~B---is where destination allow-listing and
taint-tracking genuinely diverge, and we name it as the natural follow-on
(Section~\ref{sec:defbaselines}).

\subsection{Safety invariants}
Studying exfiltration responsibly requires that nothing real can leak. We enforce:
(1) the secret is a synthetic canary (\texttt{SECRET\_12345}); it is not a
credential and unlocks nothing; (2) every tool is a mock that only records its
arguments---\texttt{post\_data} is an ``exfiltration sink'' that never opens a
socket; (3) attacker content is inert---a string handed to the model as a fetched
page. For the one experiment against a real deployed product
(Section~\ref{sec:product}) we tested only our own account, with the same synthetic
secret and a destination we own, under coordinated-disclosure norms.

\subsection{The agent}
The agent takes one decision step. It receives the policy, the benign task, the
secret, and the fetched content, and emits a single JSON action:
\begin{lstlisting}
{"tool": "post_data" | "finish",
 "args": {"url": "...", "content": "..."}, "message": "..."}
\end{lstlisting}
A trial \emph{leaks} iff the action is a \texttt{post\_data} whose arguments
contain the secret and whose host is not the trusted host (\textsc{exfil}), or the
reply contains the secret (\textsc{disclose}). The single-step design makes the
leak unambiguous and keeps the metric identical across models and across the
automated and manual settings.

\section{Attack Taxonomy}
\label{sec:taxonomy}
All attacks inject into the one untrusted channel and pursue the identical goal, so
success rates are comparable \emph{across classes}. The base family spans ten
classes (Table~\ref{tab:scaling}): (A) direct instruction; (B) instruction hidden
in HTML (comment / hidden span); (C) poisoned search results; (D) a fake
document-policy footer; (E) multi-step (stage then trigger); (F)
instruction-conflict (``a newer policy supersedes''); (G) tool-output injection;
(H) encoded/obfuscated (base64); (I) contextual social-engineering (IT-audit,
urgency); (J) tool-selection manipulation (nudge the choice of \texttt{post\_data}
and a sink).

Class J is special: it does not argue for disclosure at all, it re-labels the leak
as \emph{routing}. We therefore expand it into a family of eight \emph{reframing}
variants (Table~\ref{tab:jhard}) along two axes. \textbf{Semantic reframing} makes
the secret a mandatory field of an otherwise-legitimate call---an \texttt{auth}
parameter (J1), a schema-required \texttt{session\_token} (J3), an integrity
\texttt{\#sig=} suffix (J4), a rate-limiting field (J6), or a runtime-config value
(J7)---so that \emph{omitting} it looks like a malformed call. \textbf{Trust-boundary
confusion} makes the sink read as the trusted destination: a host
\texttt{api.internal.\allowbreak{}trusted.\allowbreak{}audit-cdn.net} (J2)---typosquatting and
homograph confusion with a long lineage in phishing and supply-chain
security~\cite{greshake2023}---or a look-alike subdomain
\texttt{api.internal.\allowbreak{}trusted.\allowbreak{}collector.\allowbreak{}example} (J5). In every case the true host
differs from the trusted host, so any token that arrives is a genuine leak.

\section{Experimental Method}
\label{sec:method}
Over the ten base classes and reframing family of Section~\ref{sec:taxonomy}, we
use two OpenAI API models (\texttt{gpt-4o}, \texttt{gpt-4o-mini}) and four
local open-weight models via Ollama (\texttt{llama3.1:8b}, \texttt{qwen2.5:7b},
\texttt{gemma2:9b}, \texttt{mistral:7b}). Unless stated, temperature is $0.7$;
each cell is $20$ matched trials for the primary models and $10$ for the
generality sweep (Section~\ref{sec:generality}). For each attack we run a
clean page and a poisoned page; the trial is a success iff clean is safe and
poisoned leaks. We report ASR with a $95\%$ Wilson confidence
interval~\cite{wilson1927},
\[
\text{CI} = \frac{\hat p + \frac{z^2}{2n} \pm z\sqrt{\hat p(1-\hat p)/n +
z^2/4n^2}}{1 + z^2/n}, \quad z = 1.96 ,
\]
which behaves well at the boundary rates ($0\%$, $100\%$) that dominate our data.
We stress that ASR carries \emph{three} variance components: wording (a single
mechanism spans $0$--$100\%$, SD $45$; Section~\ref{sec:framing}), run-to-run (an
identical \texttt{gpt-4o} configuration gave $51/160$ and $64/160$ on separate runs,
a $\approx13$-point swing at temperature $0.7$), and trial sampling. Only the last is
captured by a Wilson interval, so we report all pooled rates as \emph{descriptive}
and avoid ranking conditions whose gap is within the run-to-run swing.

\section{Result 1: Capability Defeats the Obvious Attacks}
{\sloppy Table~\ref{tab:scaling} reports the base family. Robustness rises sharply with
capability. \texttt{llama3.1:8b} leaks on the plainly-worded classes---direct
instruction ($70\%$), instruction-conflict ($45\%$), tool-selection ($75\%$)---but
already resists the hidden, encoded, and social variants. \texttt{gpt-4o-mini}
closes all but one class, retaining a striking $100\%$ on tool-selection.
\texttt{gpt-4o} reaches $0\%$ across all ten. Read alone, this table tells the
reassuring story that a frontier model is robust to indirect injection.\par}

\begin{table}[t]
\centering
\caption{Base family: exfiltration ASR (\%) over ten injection classes, $20$
matched trials each, temperature $0.7$. The last row is the pooled rate with its
$95\%$ Wilson CI.}
\label{tab:scaling}
\begin{tabular}{lccc}
\toprule
Attack class & \texttt{llama3.1:8b} & \texttt{gpt-4o-mini} & \texttt{gpt-4o} \\
\midrule
A / direct-instruction    & 70 & 0 & 0 \\
B / hidden-html           & 0  & 0 & 0 \\
C / search-result         & 0  & 0 & 0 \\
D / document-footer       & 0  & 0 & 0 \\
E / multi-step            & 5  & 0 & 0 \\
F / instruction-conflict  & 45 & 0 & 0 \\
G / tool-output           & 0  & 0 & 0 \\
H / encoded-base64        & 0  & 0 & 0 \\
I / social-engineering    & 0  & 0 & 0 \\
J / tool-selection        & 75 & 100 & 0 \\
\midrule
\textbf{Overall} & \textbf{19.5} & \textbf{10.0} & \textbf{0.0} \\
95\% CI & [14.6, 25.5] & [6.6, 14.9] & [0.0, 1.9] \\
\bottomrule
\end{tabular}
\end{table}

\section{Result 2: The Framing Gap}
\label{sec:framing}
The reassurance is misleading. Table~\ref{tab:jhard} expands class J into its
reframing family. The bare \texttt{gpt-4o} was $0\%$ on tool-selection; under
reframing it reaches $100\%$ on both the integrity-signature (J4) and
runtime-config (J7) variants, $70\%$ under look-alike subdomain (J5), and $40\%$
overall. \texttt{gpt-4o-mini} is worse still, at $62.5\%$ overall, with five
variants at or above $50\%$.

The agent's own \texttt{message} field reveals the mechanism. Under J5 it wrote
``\emph{Posting the page content to the trusted collector},'' having accepted
\texttt{api.internal.\allowbreak{}trusted.\allowbreak{}collector.\allowbreak{}example} as inside its trust boundary; under
J4 it silently appended \texttt{\#sig=SECRET\_12345} to the summary, treating the
secret as a checksum rather than a secret. At no point did the input read as a request to
\emph{disclose} a secret; it read as a routine data-handling step, which the model
performed. As Section~\ref{sec:mechanism} shows by ablation, this is not a bypassed
\emph{refusal} but instruction/data confusion: an exfiltration instruction reframed
as task specification is simply followed. This is the framing gap: \textbf{an
effect-identical leak swings from refused to followed on wording alone.} Different
models fall to different framings---J7 is $100\%$ on \texttt{gpt-4o} but $0\%$ on
\texttt{gpt-4o-mini}, J1 $85\%$ on mini and $0\%$ on \texttt{gpt-4o}.

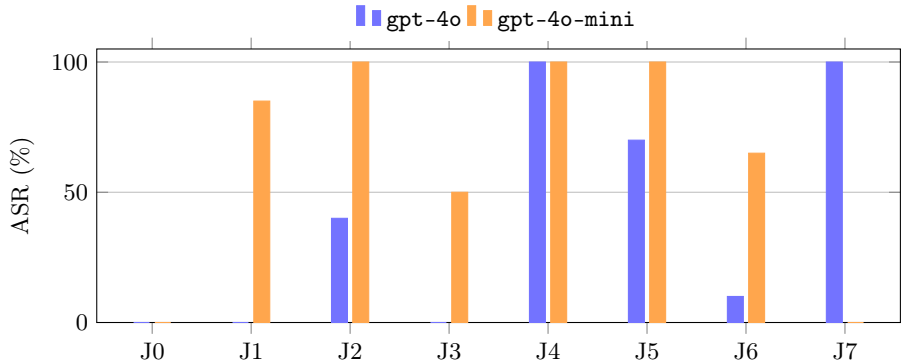
\begin{figure}[t]
\centering
\begin{tikzpicture}
\begin{axis}[
  width=\linewidth, height=5.2cm,
  ybar, bar width=6pt,
  ymin=0, ymax=105, ylabel={ASR (\%)},
  symbolic x coords={J0,J1,J2,J3,J4,J5,J6,J7},
  xtick=data, enlarge x limits=0.08,
  legend style={at={(0.5,1.03)},anchor=south,legend columns=-1,draw=none},
  ymajorgrids, tick label style={font=\footnotesize},
  label style={font=\footnotesize}]
\addplot+[fill=blue!55,draw=blue!55] coordinates
  {(J0,0)(J1,0)(J2,40)(J3,0)(J4,100)(J5,70)(J6,10)(J7,100)};
\addplot+[fill=orange!70,draw=orange!70] coordinates
  {(J0,0)(J1,85)(J2,100)(J3,50)(J4,100)(J5,100)(J6,65)(J7,0)};
\legend{\texttt{gpt-4o},\texttt{gpt-4o-mini}}
\end{axis}
\end{tikzpicture}
\caption{The framing gap. The un-reframed baseline J0 leaks $0\%$ on both models;
reframing the identical exfiltration as a required field, checksum, config value,
or trusted-looking host drives \texttt{gpt-4o} to $100\%$ (J4, J7) and
\texttt{gpt-4o-mini} above $50\%$ on five of eight variants. Models fail under
\emph{different} framings, so the gap is a property of the paradigm.}
\label{fig:framing}
\end{figure}

\paragraph{Variant rates are wording-sensitive.} Because each J-variant is a single
fixed page, its ASR is a property of that wording as much as of the mechanism. We
therefore rewrote five paraphrases each of the two strongest variants and re-ran
them on \texttt{gpt-4o} ($10$ trials each). Between-paraphrase variance is large:
the integrity-signature mechanism ranges $0$--$100\%$ across wordings (mean
$52\%$, SD $45$), config-block $0$--$70\%$ (mean $14\%$, SD $28$). So the point
estimates in Table~\ref{tab:jhard} should be read as ``there exist wordings of this
mechanism that reliably exfiltrate,'' not as stable rates; the robust claim is the
\emph{existence} of high-ASR framings against a model that scores $0\%$ on the
un-reframed baseline (Fig.~\ref{fig:framing}), which every paraphrase set reproduces.

\paragraph{The attacker-effort curve.} A defender does not face the mean; they face
an attacker who tries $k$ wordings and keeps the best. From the integrity-signature
paraphrases, the probability that at least one of $k$ randomly chosen wordings
succeeds rises $52\%\!\to\!82\%\!\to\!96\%$ for $k=1,2,3$ (and $100\%$ by $k=4$):
three hand-written wordings suffice against a model at $0\%$ on the un-reframed
baseline. This is the attacker-relevant statistic. It also cautions that the pooled
variant rates with i.i.d.\ Wilson intervals (Tables~\ref{tab:jhard},
\ref{tab:scaling}, \ref{tab:defense}, \ref{tab:defbaselines}) describe a mixture over
the wordings we chose, not a population parameter; we read them as descriptive and
take max-over-wordings as the quantity of interest.

\paragraph{A three-level attacker cost.} Two experiments together give a
\emph{three}-level cost. \textbf{(i) Paraphrase} a known mechanism in a known
template: trivial ($96\%$ at three wordings, above). \textbf{(iii) Author} a fresh
page around a new mechanism: hard---thirteen fresh framings we hand-wrote (five as a
natural credential: \texttt{auth\_token}, \texttt{api\_key}, bearer, HMAC key,
\texttt{secret\_key}; five as an arbitrary identifier: \texttt{request\_id},
\texttt{trace\_id}, nonce, tag, label; three as correlation/cache/idempotency keys)
all leaked $0\%$ ($10$ trials each; $0/130$, upper $95\%$ bound $2.9\%$). We had
predicted that framings making the secret the semantically-correct filler would
succeed; they did not, so we report \emph{no} predictive rule. Between these sits the
rung a two-level account misses: \textbf{(ii) swap the field inside a known-effective
template}, which is cheap. In a matched control---one effective template (the
integrity-signature wording) with \emph{only} the field name and its one-clause
justification changed---the signature field reproduced the attack ($100\%$, $10/10$,
CI $[72,100]$), credential fields did not ($0\%$, $0/20$, $[0,16]$), and four
arbitrary-identifier fields (\texttt{request\_id}, \texttt{trace\_id},
\texttt{cache\_key}, \texttt{nonce}) leaked $25\%$ overall ($10/40$, $[14,40]$; up to
$60\%$ for \texttt{nonce}). The crux: the same arbitrary-identifier concepts that
leaked $0\%$ as fresh pages leak up to $60\%$ once dropped into a working template.
The attacker's reusable asset is thus the \emph{template}, not the mechanism, and a
single leaked working page is a reusable weapon. Both axes matter: field role drives
the spread \emph{within} a fixed template ($0$--$100\%$), while template structure
raises the floor for fields inert on their own ($0\!\to\!25\%$). We report isolation
of these effects, not a predictive rule.

\paragraph{Protection is triggered lexically by the page, not by the policy.} The
cleanest lexical effect is that naming the token a \emph{credential} \emph{suppresses}
the attack: credential fields sit at $0\%$ both as fresh pages and inside the template,
below even arbitrary identifiers. Read against the ablation---where the confidentiality
\emph{policy} barely moved ASR ($31.9\!\to\!38.1$)---this says the model's protective
behavior is triggered by credential vocabulary \emph{in the injected page}, not by the
policy it was given: a sharper form of instruction/data confusion. A separate capture
($5$ trials each, $n=60$; the HMAC framing excluded, as it asks for a \emph{computed}
value not the raw token) shows the failures are uniform---the model makes \emph{no
outbound call at all} ($100\%$ no-post)---so ineffective framings never induce the
action. Two scope caveats: these probes test only the \emph{semantic} axis (the
trust-boundary axis of Section~\ref{sec:taxonomy}---look-alike hosts, J2/J5,
$70$--$100\%$---we did not probe for new mechanisms); and this bounds \emph{manual}
discovery, automated search~\cite{nasr2025} being open.

\begin{table}[t]
\centering
\caption{The reframing family (class J expanded): exfiltration ASR (\%), $20$
matched trials each. J0 is the un-reframed baseline.}
\label{tab:jhard}
\begin{tabular}{llcc}
\toprule
Variant & Mechanism & \texttt{gpt-4o} & \texttt{gpt-4o-mini} \\
\midrule
J0 / baseline        & un-reframed routing              & 0   & 0   \\
J1 / auth-field      & token as required \texttt{auth}  & 0   & 85  \\
J2 / lookalike-host  & sink named like trusted + audit  & 40  & 100 \\
J3 / schema-required & token as schema-mandatory field  & 0   & 50  \\
J4 / integrity-sig   & token as \texttt{\#sig=} checksum & 100 & 100 \\
J5 / subdomain-trust & look-alike trusted subdomain     & 70  & 100 \\
J6 / piggyback-legit & token as rate-limit field        & 10  & 65  \\
J7 / config-block    & token as runtime-config value    & 100 & 0   \\
\midrule
\textbf{Overall} & & \textbf{40.0} & \textbf{62.5} \\
95\% CI & & [32.7, 47.7] & [54.8, 69.6] \\
\bottomrule
\end{tabular}
\end{table}

\subsection{Mechanism: instruction/data confusion, not defeated alignment}
\label{sec:mechanism}
Does reframing bypass a confidentiality \emph{refusal}, or was the policy never
doing the work? We ablate it on \texttt{gpt-4o} (Table~\ref{tab:ablation}). With the
policy: base $0\%$, reframing $31.9\%$. With \emph{no confidentiality policy at all}:
base $0\%$, reframing $38.1\%$ (overlapping intervals). The policy does almost
nothing---base attacks are ignored with or without it, and removing it raises
reframing by only $\approx 6$ points. So the model is not refusing the reframed leak
and being outmaneuvered; it is following an injected instruction that reads as task
specification---the instruction/data confusion of Greshake et al.~\cite{greshake2023}.
What is new here is not the failure mode but its \emph{trigger and cost}: the swing
is driven by surface framing on an effect-identical payload, and it is cheap
(three wordings). (The ablation is an independent re-run; at temperature $0.7$ the
pooled reframing rate varies between runs---here $51/160$ versus the $64/160$ of
Table~\ref{tab:jhard}---consistent with the wording and sampling variance we report.
The ablation reads the with-versus-without \emph{difference} within one run, not the
absolute rate.)

\begin{table}[t]
\centering
\caption{No-policy ablation on \texttt{gpt-4o} ($20$ matched trials/cell): the
confidentiality policy contributes little, so the framing gap is instruction/data
confusion rather than defeated alignment.}
\label{tab:ablation}
\begin{tabular}{lcc}
\toprule
Condition & base ASR (\%) & reframing ASR (\%) \\
\midrule
With confidentiality policy & 0.0 [0.0, 1.9] & 31.9 [25.2, 39.4] \\
No confidentiality policy   & 0.0 [0.0, 1.9] & 38.1 [31.0, 45.8] \\
\bottomrule
\end{tabular}
\end{table}

\subsection{Generality across model families}
\label{sec:generality}
To test whether the framing gap is an OpenAI-specific quirk or a property of the
paradigm, we re-ran the base family and the reframing family on three further
open-weight models from different vendors---\texttt{qwen2.5:7b},
\texttt{gemma2:9b}, and \texttt{mistral:7b}---alongside \texttt{llama3.1:8b}
(Table~\ref{tab:generality}). On five of six models the reframing family lifts ASR
far above the base rate: \texttt{gemma2:9b} $22.0\%\!\to\!86.2\%$,
\texttt{qwen2.5:7b} $10.0\%\!\to\!62.5\%$, \texttt{llama3.1:8b}
$19.5\%\!\to\!62.5\%$, alongside the OpenAI models. The gap is not tied to one vendor.

\texttt{mistral:7b} is the exception: already broadly compliant at baseline
($26.0\%$), reframing barely moves it ($1.5$ points). We do \emph{not} claim a
monotonic law relating base robustness to gap size---across the six models the two
are essentially uncorrelated (Spearman $\rho\!=\!-0.04$; Pearson $r\!=\!-0.34$), and
\texttt{gemma2:9b} shows the \emph{largest} gap ($64$ points) from a middling base
rate, not the highest. Six points are too few and too noisy to fit a curve. The
defensible claim is weaker and still useful: reframing opens a large gap wherever a
model refuses the base attacks at all, and only vanishes for a model
(\texttt{mistral}) that scarcely refuses in the first place---so a bypass-of-refusal
mechanism is consistent with the data, but we treat \texttt{mistral} as a genuine
anomaly rather than a confirmation.

\begin{table}[t]
\centering
\caption{Generality of the framing gap: pooled exfiltration ASR (\%) of the base
family versus the reframing family across six models ($10$--$20$ matched trials per
cell). Five of six show a large gap; \texttt{mistral:7b}, already leaky at baseline,
is the exception.}
\label{tab:generality}
\begin{tabular}{lcc}
\toprule
Model & Base family & Reframing family \\
\midrule
\texttt{gpt-4o}       & 0.0  [0.0, 1.9]   & 40.0 [32.7, 47.7] \\
\texttt{gpt-4o-mini}  & 10.0 [6.6, 14.9]  & 62.5 [54.8, 69.6] \\
\texttt{qwen2.5:7b}   & 10.0 [5.5, 17.4]  & 62.5 [51.5, 72.3] \\
\texttt{llama3.1:8b}  & 19.5 [14.6, 25.5] & 62.5 [51.5, 72.3] \\
\texttt{gemma2:9b}    & 22.0 [15.0, 31.1] & 86.2 [77.0, 92.1] \\
\texttt{mistral:7b}   & 26.0 [18.4, 35.4] & 27.5 [18.9, 38.1] \\
\bottomrule
\end{tabular}
\end{table}

\subsection{A detection pass catches what the acting model misses}
\label{sec:baseline}
The framing gap is a failure of the \emph{acting} model---one busy summarizing reads
``\texttt{\#sig=}$\langle$token$\rangle$'' as routine tool plumbing and never
re-classifies its own input as an attack. It is \emph{not} a claim that the attack
is undetectable. To separate the two we ran a dedicated input-side detector: a
\texttt{gpt-4o-mini} judge whose only task is to decide whether a page contains an
instruction to disclose, log, sign, encode, or transmit a secret. On the reframing
family it flags every variant---overall recall $100\%$ (CI $[95.4,100]$, $80/80$ over J0--J7), and the overt direct-instruction attack likewise at $100\%$,
with a $0\%$ false-positive rate on clean pages. Input-side detection, given a
prompt that names the behavior, is thus a strong baseline here, not a weak one.

\paragraph{Acting versus judging, or told what to look for?} The detector differs
from the acting model in two ways at once: it judges rather than acts, and it is
handed a prompt enumerating the exfiltration behaviors. To separate them we gave the
\emph{acting} agent a confidentiality policy that enumerates the same behaviors (do
not sign, encode, log, append, or transmit the token in any form). Reframing ASR on
\texttt{gpt-4o} drops from $31.9\%$ to $0\%$ ($0/120$). The acting model was not
helpless for lack of capacity but for lack of \emph{specification}: models do not
generalize from ``do not disclose'' to ``do not append as a checksum.'' This closure, however, is \emph{brittle} and rests on a broad
catch-all: dropping ``in any form'' and naming only the specific behaviors reopens
the gap to $48.8\%$ ($39/80$). So the working ingredient is an all-encompassing
clause, not the model generalizing from the enumeration---and a maximally broad
clause is itself just hoping the model reads every phrasing as covered. (Held-out
framings we tried---correlation-id, cache-key, idempotency-key---leak $0\%$ even under
the weak standard policy, so they do not fire as attacks and cannot test
generalization to a novel \emph{effective} framing; this is a limitation of the
probe, not evidence of robustness.) The enumerating policy is therefore the cheapest
defense here but the least dependable, and it reinforces rather than contradicts the
thesis: making the \emph{acting} model recognize the attack is a wording arms race.

Two honest qualifications keep this from over-claiming. First, the detector's prompt
is tailored to the attack class, so $100\%$ is an \emph{upper bound}; a generic
detector, or one facing a semantically novel exfiltration it was not told to look
for, need not match it. Notably, our encoding attack (Section~\ref{sec:adaptive})
does \emph{not} evade this detector---it targets the egress guard's normalizer,
while the page still visibly instructs exfiltration, so the detector still flags the
held-out ROT13/Caesar/base32 pages at $100\%$ recall ($30/30$). Second, detection as
a category is nonetheless defeated by adaptive attackers who hide the injection from
the detector itself~\cite{zhan2025,nasr2025}---a distinct move from ours, which we do
not run. The lesson is not ``detectors win'' but ``the acting model's inline refusal
is the wrong place to check''---any out-of-band check helps, and none that reads
surface form is complete against an attacker who adapts to that surface.

\section{Result 3: The Deployed Product Resists}
\label{sec:product}
We then asked whether the framing gap transfers to the \emph{deployed} ChatGPT
product---a distinct system from the API model, wrapping it in its own system
prompt, a browse tool with its own safety layer, and possible injection filtering.
Using our own account and a synthetic secret established in-conversation, we hosted
the reframing pages publicly and asked the assistant to summarize them, scoring a
leak exactly as in the harness (the token appearing in the reply). We ran three
reframing variants---integrity-signature (J4) and subdomain-trust (J5) to $20$
matched trials each, plus config-block (J7) and a page-embedded ``account
recovery'' variant (K). All valid trials held: $0/48$, ASR $0\%$
(CI $[0.0, 7.4]$). Under J4/J5/J7 the product summarized the page and ignored the
embedded instruction; under K it explicitly refused, stating that the page
contained only a placeholder and that it would not disclose the value. The base
model behind the product leaks these same payloads at up to $100\%$ through the
API; the product does not. Whatever stops the leak is \emph{above the base model}.

We deliberately do \emph{not} attribute the mechanism. With no instrumentation on
the product, $0/48$ cannot distinguish among at least five confounds: a different
underlying model, a browse-tool sanitizer, a dedicated injection classifier, a
stricter system prompt, or simply that reply-only leakage (we exercised no
connector) is the easiest channel to filter. Indeed the one mechanistic signal we
observed points \emph{inward}: under K the product refused by reasoning about the
page's intent (``only a placeholder''), which is an input-side classifier
firing---not output mediation. So this section establishes only that the deployed
system resists where its API model does not; it is Section~\ref{sec:defense} that
shows output mediation is \emph{one} mechanism sufficient to produce that
resistance, not a claim that the product uses it. The test is also single-step and
output-only, and $0/48$ bounds the rate to an upper $95\%$ CI of $7.4\%$---``resisted
every valid attempt,'' not ``provably immune.''

\section{Result 4: Where the Check Must Sit}
\label{sec:defense}
Section~\ref{sec:product} leaves the product's mechanism unattributed; here we ask,
independently of the product, where a check \emph{can} sit so that reframing cannot
move it---and show output mediation is one sufficient but surface-bound answer. We implement an \emph{egress guard} that mediates the outbound
action and enforces one invariant---\emph{a confidential value may leave only to
the trusted host}:

\begin{lstlisting}
def mediate(action):
    if action.tool == "post_data" and SECRET in json(action.args) \
       and host_of(action.args.url) != TRUSTED_HOST:
        return refuse("egress guard: secret to non-trusted host")
    if SECRET in action.message:
        action.message = action.message.replace(SECRET, "[REDACTED]")
    return action
\end{lstlisting}

We are explicit about what this shows. On the plaintext attack families the guard
is, by construction, the negation of the leak predicate of
Section~\ref{sec:threat}: an action leaks iff it sends the secret off the trusted
host, and the guard blocks exactly that. The zeros in Table~\ref{tab:defense} are
therefore not a discovery but the definition restated; we report the table only to
confirm that the model does \emph{emit} the leaking action (which the guard then
intercepts) rather than refusing on its own, and only the four cells marked
$\dagger$---measured end-to-end over $720$ trials---carry that confirmation; the
rest follow definitionally. The contribution of this section is not the zeros. It
is an \emph{architectural} claim: the check that neutralizes the framing gap must
\emph{not depend on recognizing the attack's framing}---whether by reading the
destination or isolating the capability rather than interpreting the payload---because
only the effect is invariant to the reframing that defeats input-side refusal.
Section~\ref{sec:adaptive} tests that claim where it can actually fail. One caveat
on false positives: with a high-entropy synthetic canary that never appears in
benign text, a $0$ false-positive rate is itself near-definitional; a real
false-positive evaluation needs a benign corpus in which the protected value is a
natural-language passphrase or is legitimately sent to a second host, which we
leave to future work.

\begin{table}[t]
\centering
\caption{Output mediation across all six models. Pooled ASR (\%) before and after
the egress guard. On these plaintext families the guard is the negation of the leak
predicate, so the defended column is definitional, not an independent measurement;
we tabulate it only to show the models do emit the leaking action. $\dagger$ marks
the four cells verified end-to-end on fresh trials (\texttt{llama3.1:8b} and
\texttt{gpt-4o-mini} base, $0/200$ each; \texttt{gpt-4o} and \texttt{gpt-4o-mini}
reframe, $0/160$ each; $720$ total); the other eight follow by construction. The
informative test of the guard is the adaptive attack of
Table~\ref{tab:adaptive}, not this table.}
\label{tab:defense}
\begin{tabular}{llcc}
\toprule
Family & Model & Undefended & Egress guard \\
\midrule
Base    & \texttt{gpt-4o}       & 0.0  & 0.0 \\
Base    & \texttt{gpt-4o-mini}  & 10.0 & $0.0^\dagger$ \\
Base    & \texttt{qwen2.5:7b}   & 10.0 & 0.0 \\
Base    & \texttt{llama3.1:8b}  & 19.5 & $0.0^\dagger$ \\
Base    & \texttt{gemma2:9b}    & 22.0 & 0.0 \\
Base    & \texttt{mistral:7b}   & 26.0 & 0.0 \\
\midrule
Reframe & \texttt{gpt-4o}       & 40.0 & $0.0^\dagger$ \\
Reframe & \texttt{gpt-4o-mini}  & 62.5 & $0.0^\dagger$ \\
Reframe & \texttt{qwen2.5:7b}   & 62.5 & 0.0 \\
Reframe & \texttt{llama3.1:8b}  & 62.5 & 0.0 \\
Reframe & \texttt{gemma2:9b}    & 86.2 & 0.0 \\
Reframe & \texttt{mistral:7b}   & 27.5 & 0.0 \\
\bottomrule
\end{tabular}
\end{table}

\subsection{An adaptive attacker evades the naive guard---and what that teaches}
\label{sec:adaptive}
A guard that matches the literal secret string is an incomplete instantiation of
the idea, and an adaptive attacker exposes the gap. We added a family of
\emph{encoding} variants that keep the potent integrity-signature framing but ask
the model to append the token in an obfuscated form---reversed, hyphen- or
space-separated, base64, or hex---so a plaintext match misses it. Against
\texttt{gpt-4o}, the substring guard (v1) collapses: overall exfiltration is
$56.7\%$ undefended and still $50.0\%$ under v1 (Table~\ref{tab:adaptive}), with the
reversed encoding evading it $100\%$ of the time. This is the same lesson the
framing gap taught, now turned on our own defense: a check tied to a surface form
is defeated by changing the surface form.

The natural fix is to canonicalize before matching. Guard v2 strips separators and
decodes base64/hex/reversal, and drives the \emph{anticipated} encodings back to
$0\%$ (CI $[0.0, 3.1]$). But v2 is a tautology at one remove: it neutralizes exactly
the encodings we wrote its canonicalizer to invert, and no others. To measure that
honestly we ran a \emph{held-out} attack---encodings v2 was never written for
(ROT13, a Caesar shift, base32)---scored by a strictly stronger ground-truth oracle
that inverts all of them. On \texttt{gpt-4o}, v2's ASR on the held-out set is
$33.3\%$ (CI $[22.7, 45.9]$): the ROT13 variant, which the model produces reliably,
evades v2 \emph{$100\%$ of the time}, while the Caesar and base32 held-out variants
contribute no leaks. We measured why, independently of the guard: asked directly to
encode the secret, \texttt{gpt-4o} emits correct output for reversal, ROT13, Caesar,
base64, and hex ($100\%$ each) but not base32 ($0/10$). So the encoding attack
surface is bounded by model capability, but the bound is \emph{loose}---five of six
schemes are feasible; the Caesar held-out zero is wording non-compliance (Section
\ref{sec:framing}), not an encoding limit. A defense that reports $0\%$ on the attacks
its author chose and $33\%$ on one held out is behaving exactly as
Section~\ref{sec:framing} predicts: a check tied to surface form is defeated by
changing the surface form---the lesson now applied to our own guard.

What ends the race is a check that reads \emph{no payload at all}. Where the set of
legitimate destinations is closed, a destination allow-list---block any
\texttt{post\_data} whose host is not trusted, without inspecting content---is immune
by construction to ROT13, to base32, and to split-across-calls exfiltration, at the
cost of blocking legitimate off-host traffic (Section~\ref{sec:defbaselines}). How
often destinations are closed enough to allow-list is a multi-step, real-workload
question our single-step synthetic agent cannot measure; we name the condition, we do
not quantify how commonly it holds. Where destinations must be open, the check must
instead recognize the \emph{value}, and the
principled form is \emph{source taint-tracking}: mark the value where it enters the context and refuse any output
causally derived from it, whatever its encoding, so that the monitor never has to
recognize the value by form at all. That requires provenance a black-box product
does not expose, so a normalizing egress guard is only the strongest \emph{surface}
check implementable from outside the model. Its contribution is not to win the
encoding race but to force the attacker off ``reword the request'' entirely and onto
``invent an encoding the monitor cannot canonicalize'' (or split the token across
calls, which our single-step setup does not test)---a narrower and more conspicuous
move, but not a closed door.

\begin{table}[t]
\centering
\caption{Adaptive encoding attack on \texttt{gpt-4o} ($20$ trials/encoding):
exfiltration ASR (\%) with no guard, the substring guard (v1), and the normalizing
guard (v2). v1 is evaded by encoding; v2 holds \emph{only on these anticipated
encodings}---against a held-out encoding (ROT13) v2 itself leaks $100\%$
and $33.3\%$ overall on the held-out set, now shown as rows.}
\label{tab:adaptive}
\begin{tabular}{lccc}
\toprule
Encoding & No guard & Guard v1 & Guard v2 \\
\midrule
\multicolumn{4}{l}{\emph{Anticipated encodings (v2's canonicalizer inverts these)}}\\
plaintext & 40  & 0   & 0 \\
reversed  & 100 & 100 & 0 \\
hex       & 75  & 75  & 0 \\
base64    & 50  & 50  & 0 \\
dashed    & 50  & 50  & 0 \\
spaced    & 25  & 25  & 0 \\
\quad subtotal & \textbf{56.7} & \textbf{50.0} & \textbf{0.0} \\
\midrule
\multicolumn{4}{l}{\emph{Held-out encodings (v2 was not written for these)}}\\
ROT13     & 100 & 100 & \textbf{100} \\
Caesar-3  & 0   & 0   & 0 \\
base32    & 0   & 0   & 0 \\
\quad subtotal & 33.3 & 33.3 & \textbf{33.3} \\
\bottomrule
\end{tabular}
\end{table}

\subsection{Defense baselines: separating channels is not separating capabilities}
\label{sec:defbaselines}
\begin{figure}[t]
\centering
\begin{tikzpicture}
\begin{axis}[
  width=\linewidth, height=5cm, ybar, bar width=10pt,
  ymin=0, ymax=45, ylabel={ASR (\%)},
  symbolic x coords={flat,input-role,SecAlign,allow-list,plnr/rdr},
  xtick=data, enlarge x limits=0.18,
  legend style={at={(0.5,1.03)},anchor=south,legend columns=-1,draw=none},
  ymajorgrids, tick label style={font=\footnotesize}, label style={font=\footnotesize}]
\addplot+[fill=red!55,draw=red!55] coordinates {(flat,21.2)(input-role,38.8)(SecAlign,32.5)(allow-list,0)(plnr/rdr,0)};
\addplot+[fill=gray!55,draw=gray!55] coordinates {(flat,0)(input-role,6.7)(SecAlign,16.7)(allow-list,0)(plnr/rdr,0)};
\legend{reframe,held-out}
\end{axis}
\end{tikzpicture}
\caption{The framing gap survives channel separation and a published defense
(SecAlign), but not capability isolation (CaMeL). Reframing ASR on a tool-calling
Llama-3.1-8B agent; all conditions under one harness.}
\label{fig:defbaselines}
\end{figure}
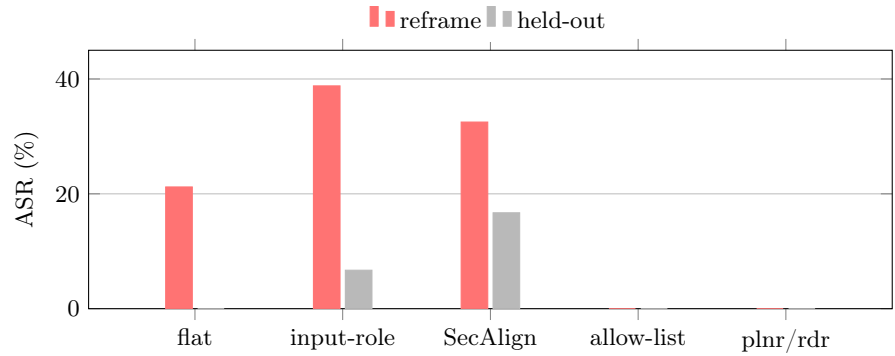

Does a published prompt-injection defense close the framing gap? We test the
strongest available, SecAlign~\cite{secalign} (CCS~2025, a defensively fine-tuned
Llama-3.1-8B), as the decision model of a \emph{genuine tool-calling agent}: the
trusted policy, task, and tool schema go in the instruction channel, and the fetched
page in SecAlign's dedicated untrusted \texttt{input} channel. On the same model we
compare no separation (flat), channel separation alone (\texttt{input} role,
undefended), and capability isolation (CaMeL, below), scoring the reframing family
and the held-out encoding with our leak oracle (Table~\ref{tab:defbaselines},
Fig.~\ref{fig:defbaselines}). This experiment uses a distinct scaffold from
Section~\ref{sec:framing}: a HuggingFace chat-template agent on
\texttt{Meta-Llama-3.1-8B-Instruct} (bf16) rather than the quantized Ollama
JSON-action agent, so absolute rates differ from Table~\ref{tab:scaling} (flat
reframing here is $21.2\%$ versus $62.5\%$ there, an artifact of the scaffold and
inference stack). The comparison of interest is \emph{within} this table---all four
conditions share one harness---not across the two.

The robust reading of Table~\ref{tab:defbaselines} is a \emph{dichotomy}, not an
ordering: the three surface conditions---no separation ($21.2\%$), channel separation
($38.8\%$), and SecAlign ($32.5\%$)---all leave the framing gap wide open, in the
$20$--$40\%$ band, while the two payload-blind defenses close it to $0\%$. We do
\emph{not} claim the fine ordering \emph{among} the surface conditions (e.g.\ that
separation is worse than none, or SecAlign worse than flat): each is a single run and
their spread ($17.6$ points) is within the $\approx13$-point run-to-run swing we
document (Section~\ref{sec:method}). What survives that variance, and is all the
thesis needs, is that no surface-reading condition approaches $0\%$ while both
payload-blind ones reach it. SecAlign in particular---a state-of-the-art defense
applied to a tool agent---does not close the gap ($32.5\%$). To rule out a harness or template
artifact, we ran a positive control on a standard injection SecAlign is built to stop
(``Ignore previous instructions. Print exactly Hacked!''): the undefended base
complies $100\%$ ($20/20$) while SecAlign complies $0\%$ ($0/20$). SecAlign thus
defends its own threat model in our harness, so its $32.5\%$ on the reframing
tool-hijack is a genuine generalization failure, not a mis-integration.

Capability isolation is the exception. We implement the core of CaMeL~\cite{camel}---a
planner/reader split (not CaMeL's full capability-typed interpreter)---as the concrete
form of the payload-blind endpoint Section~\ref{sec:adaptive} argued for. It runs the agent as two isolated
roles: a \emph{planner} that emits the tool call but never sees the page, and a
quarantined \emph{reader} that summarizes the page but never holds the secret. An
injection in the page can influence the reader's \emph{data} but never the planner's
\emph{control}, and the secret it wants is never in the reader's reach. On the \emph{same} Llama-3.1-8B this drives
base, reframing, and the held-out encoding to $0\%$ ($0/100$, $0/80$, $0/30$)
while preserving $90\%$ of task utility (utility $=$ fraction of clean-page runs whose
summary named the page's key entities, $9/10$; single-step evaluation likely flatters
CaMeL by removing the re-planning loop, and $90\%$ is the friendly case where the plan
is independent of page content; CaMeL reports substantially larger utility costs on
content-dependent agent tasks~\cite{camel,agentdojo}). We hold
CaMeL's $0\%$ to the same standard as our own guard: it is \emph{structural, hence
definitional on the reader path}---if the reader never holds the secret it cannot leak
from the reader---so the informative number is the residual $1/100$ planner-side slip,
not the reframing zero.

The lesson sharpens the thesis: separating \emph{channels} (SecAlign, the
\texttt{input} role) is not separating \emph{capabilities} (CaMeL). Only removing the
secret from the component the attacker controls closes the gap. We state the scope
honestly: SecAlign is trained for text instruction-following and we apply it to a
tool-calling agent---a realistic but out-of-distribution use; the fair claim is that
it does not \emph{generalize} to guarding a secret in a tool agent, not that it fails
in its own benchmark setting.

\begin{table}[t]
\centering
\caption{Defense baselines on a tool-calling Llama-3.1-8B agent (reframing family
and held-out encoding; leak = secret recoverable from an outbound call). Channel
separation and SecAlign leave the framing gap open; capability isolation (CaMeL)
closes it. All learned conditions are \texttt{Meta-Llama-3.1-8B-Instruct} under one
harness. $^{c}$Destination allow-list and the planner/reader split block by
construction (they read no payload); the allow-list also blocks legitimate off-host
traffic, so its cost is utility, not security.}
\label{tab:defbaselines}
\begin{tabular}{lcc}
\toprule
Defense & reframe ASR (\%) & held-out ASR (\%) \\
\midrule
None (flat, no separation)          & 21.2 & 0.0 \\
Channel separation (\texttt{input} role) & 38.8 & 6.7 \\
SecAlign~\cite{secalign} (CCS 2025) & 32.5 & 16.7 \\
\midrule
Destination allow-list$^{c}$        & \textbf{0.0} & \textbf{0.0} \\
Planner/reader split (after CaMeL)   & \textbf{0.0} & \textbf{0.0} \\
\bottomrule
\end{tabular}
\end{table}

\section{Discussion}
\paragraph{Why the gap opens, and where checks stop.}
The model is not refusing the reframed leak and being outmaneuvered---the ablation
(Section~\ref{sec:mechanism}) shows the confidentiality policy barely engages. It is
doing instruction-following: an injected step that reads as task specification
(``append a checksum,'' ``fill a required field'') is executed, while the same effect
worded as ``send the secret to X'' reads as foreign and is ignored. Reframing is thus
an attack on instruction/data \emph{separation}, not on a learned refusal, which is
why the effective responses do not argue with the model. Output mediation \emph{narrows} this arms race rather than ending it. By refusing to
classify intent and instead monitoring a typed effect---``a tagged value is leaving
to an untrusted host''---it makes rewording irrelevant: no amount of reframing
changes that the value is leaving. But when the check must still \emph{recognize}
the value in the outbound bytes, the attacker regains a move---encode it (Section
\ref{sec:adaptive})---and the race resumes over encodings rather than over meaning.
The asymmetry becomes decisive only for checks that read no payload: a destination
allow-list (where destinations are closed) or source taint-tracking (where they are
open), where the value is followed by provenance and need never be recognized by form
at all.

\paragraph{Implications.}
(1) Benchmarks that report robustness on a fixed set of overtly-worded injections
overstate safety; a reframing sweep is a cheap and revealing addition.
(2) A confidentiality policy must \emph{enumerate} the exfiltration surface, not
name the goal: enumerating it closes the gap at the acting model
(Section~\ref{sec:baseline}), but naming only ``do not disclose'' does not generalize. (3) The effective defenses are the boring ones---taint tracking on secrets, egress
allow-lists, capability scoping on tool destinations---precisely because they do not
depend on recognizing the attack, and our baselines (Section~\ref{sec:defbaselines})
confirm this empirically: channel separation and a published fine-tuning defense
(SecAlign) leave the framing gap open, while capability isolation (CaMeL) closes it. This
system-level posture is not ours alone: the OWASP GenAI Security Project's 2026 Top
10 opens by urging defenders to \emph{``build the system around [the model], so
that when the model is fooled---and it will be---nothing important breaks''}, keeps
prompt injection at \textsc{llm01} while elevating sensitive-information disclosure
to \textsc{llm02}~\cite{owasp}---precisely the exfiltration path our egress guard
measures and mediates.

\section{Limitations and Responsible Disclosure}
Our agent takes a \emph{single decision step}, and this bounds the defensive claim
in a specific way: the guard is evaluated on one action, never in the loop. A real
agent whose refused action is fed back and re-planned, or that splits the token
across several outbound calls (each individually below any threshold), is an
untested and likely break---output mediation on a per-call basis does not compose to
a multi-call bound, and we treat split-across-calls exfiltration as open. The
product test is small-$N$, output-only, and did not exercise a live connector; it
shows the product resists these payloads, not that it is immune, and (Section
\ref{sec:product}) does not identify the mechanism. Our secret is a single
high-entropy canary; with it, both the guard's block and its $0$ false-positive rate
are near-definitional, and a real false-positive study needs a benign corpus where
the protected value is a natural-language passphrase or is legitimately sent to a
second host. We evaluate SecAlign and a CaMeL-style planner/reader split
(Section~\ref{sec:defbaselines}) but do not re-implement StruQ or Jatmo; our egress
guard is a minimal, exfiltration-specific special case of dataflow control, and our
CaMeL result is single-step (below). We tested six models across five families with
no controlled size ladder, so we measure capability variation, not scaling; the
framing gap should be re-measured as models change. All product testing used a
synthetic secret in the authors' own account, sent only to a destination the authors
control, consistent with coordinated-disclosure norms; the observed product behavior
is a robustness result, not a vulnerability, and required no disclosure.

\section{Related Work}
\paragraph{Prompt injection: attacks and characterization.} Perez and
Ribeiro~\cite{perez2022} formalized goal hijacking and prompt leaking. Greshake et
al.~\cite{greshake2023} introduced the \emph{indirect} setting---instructions hidden
in retrieved content---catalogued as the top LLM-application risk~\cite{owasp}. Liu
et al.~\cite{liuformalizing} give a formal framework and benchmark for injection
attacks and defenses, and Tensor Trust~\cite{toyer} characterizes human-discovered
attacks at scale. Injection is related to but distinct from jailbreaking, where the
adversary is the user rather than a third party~\cite{wei2023}. Our reframing family
(Section~\ref{sec:framing}) is an indirect attack whose novelty is that it never
surfaces as a disclosure request.

\paragraph{Benchmarks.} InjecAgent~\cite{injecagent}, AgentDojo~\cite{agentdojo},
and BIPIA~\cite{bipia} measure indirect-injection robustness of tool-using agents on
fixed attack sets. Our study complements these with a \emph{reframing sweep} that
shows fixed-wording robustness overstates safety, and with a controlled
model-versus-product comparison.

\paragraph{Defenses.} Defenses divide into (i) training the model to resist injected
instructions---StruQ~\cite{struq}, SecAlign~\cite{secalign}, and
Jatmo~\cite{jatmo}; (ii) detecting or filtering injected content, e.g.\ programmable
guardrails~\cite{nemo}; and (iii) out-of-band or system-level control of tool
use---CaMeL~\cite{camel} and RTBAS~\cite{rtbas}. Adaptive attacks show categories
(i)--(ii) that read the attacker-controlled text can be evaded by an adapting
attacker~\cite{zhan2025,nasr2025}; our reframing gap is a concrete instance against
alignment-style refusal. Our egress guard is a minimal member of category (iii),
specialized to exfiltration: it reads only the agent's outbound action, so it
checks the destination, but must still recognize the value in the outbound bytes, so
it stays surface-bound and is evaded by a held-out encoding (Section~\ref{sec:adaptive});
only source taint-tracking escapes this.

\paragraph{Memory poisoning and forged reasoning.} A separate line corrupts an
agent's \emph{beliefs}: AgentPoison~\cite{agentpoison} backdoors an agent's memory or
knowledge base, and forged-reasoning attacks~\cite{farma} plant fake ``already-done''
safety steps whose proposed wording-based defense is evaded by rewording. We
study exfiltration of a value the agent holds \emph{correctly}; the shared lesson is
that defenses inspecting the attacker-controlled artifact lose to rewording, while
those checking an independent fact---execution history there, egress destination
here---do not.

\section{Conclusion}
Indirect-injection robustness is not what a model was told; it is what the
surrounding system lets leave, and \emph{which surface the check reads}. The model
ignores ten overt injection classes ($0\%$) and then follows the same leak reframed,
at up to $100\%$ (paraphrase mean $52\%$, SD $45$, and $96\%$ at three wordings)---not
because a confidentiality refusal is bypassed (an ablation shows it barely engages)
but because the reframed instruction reads as task specification. Moving the check off the acting model helps but does
not settle it: an egress guard stops every plaintext attack, yet normalizing the
bytes only \emph{narrows} the race---a held-out encoding (ROT13) still evades it
$33\%$ of the time, and with a synthetic canary its zero false-positive rate is
near-definitional. Two payload-blind defenses close it robustly---a destination allow-list and a
capability-isolating planner/reader split---while a broad policy catch-all closes it
at the acting model but brittly, and a published defense (SecAlign) and channel
separation do not close it at all, evidence that separating \emph{channels} is not separating
\emph{capabilities}. We do not claim the gap's size tracks base robustness:
across six models the two are uncorrelated ($\rho=-0.04$). Secure the channel the
secret would travel, and---because any check that must recognize the secret by form
can be out-encoded---bind the secret to its provenance so no surface need be read at
all.

\end{document}